# Influence of laser spot size at diffuser plane on the longitudinal spatial coherence function of optical coherence microscopy system

Kashif Usmani[1], Azeem Ahmad[2], Rakesh Joshi[1], Vishesh Dubey[1], Ankit Butola[1], Dalip Singh Mehta[1]

[1]*Department of Physics, Indian Institute of Technology Delhi, Hauz Khas, New Delhi 110016, India*
[2]*Department of Physics and Technology, UiT The Arctic University of Norway, Tromsø 9037, Norway*
*Corresponding author:* [3]*ahmadazeem870@gmail.com*, [4]*mehtads@physics.iitd.ac.in*

**Abstract:**

Coherence properties and wavelength of light sources are indispensable for optical coherence microscopy/tomography (OCM/T) as they greatly influence the signal to noise ratio, axial resolution, and penetration depth of the system. In the present letter, we investigated the longitudinal spatial coherence properties of pseudo/thermal light source (PTS) as a function of spot size at the diffuser plane, which is controlled by translating microscope objective lens towards or away from the diffuser plane. The axial resolution of PTS is found to be maximum ~ 13 μm for the beam spot size of 3.5 mm at the diffuser plane. The change in the axial resolution of the system as the spot size is increased at the diffuser plane is further confirmed by performing experiments on standard gauge blocks of height difference of 15 μm. Thus, by appropriately choosing the beam spot size at the diffuser plane, any monochromatic laser light source depending on the biological window can be utilized to obtain high axial-resolution with large penetration depth and speckle free tomographic images of multilayered biological specimens irrespective of the source's temporal coherence length. In addition, PTS could be an attractive alternative light source for achieving high axial-resolution without needing chromatic aberration corrected optics and dispersion-compensation mechanism unlike conventional setups.

1. Introduction

Coherence properties of laser made a great impact in the field of various optical interferometry, digital holography and optical coherence microscopy/tomography (OCM/T)[1-4]. Lasers being high temporal coherence (TC) in nature can generate interference pattern quickly in optical interference microscopy systems, which further leads to the measurement of various parameters like phase/height map, refractive index and dry mass density of the biological specimens[5]. However, it reduces the interferogram's quality significantly due to the formation of speckle noise, coherent noise and parasitic fringe formation[5,6]. Moreover, high TC length of the lasers makes them unsuitable for OCM and OCT applications. In OCM/T, low TC light sources such as halogen lamp, LEDs and broadband lasers are generally employed to obtain interference pattern only from a selected planes[1,2]. Employment of low TC light source is advantageous for coherent artifacts free phase and optically sectioned images of the biological specimens, however, these sources severely suffer from the problems of chromatic aberration and dispersion[7,8].

Recently, a number of studies have been put forward to synthesize a spatially extended and temporally coherent light source named pseudo-thermal light source (PTS) to overcome aforementioned issues with the conventional light sources[3,7,9,10]. PTS can be generated by employing different approaches such as rotating diffuser, vibrating multiple multi-mode fiber bundle (MMFB) and electro-active-polymer rotational micro-optic diffuser[7,11-13]. The synthesized light carries the advantages of both high TC and low TC light sources. This makes PTS an attractive alternative light source for industrial and biological applications. PTS have been successfully implemented in optical profilometry, quantitative phase microscopy and OCM/T techniques to improve their performance[3,7,9]. In the past, a lot of theoretical and experimental work has also been done to understand and study the coherence properties of PTS. The coherence properties, especially the longitudinal spatial coherence (LSC) properties, of PTS as a function of source size, numerical aperture (NA) of microscope objective and wavelength have been studied previously[9,11,12,14,15].

It becomes important to achieve short LSC length for high resolution optical sectioning of the biological specimens. In the present work, PTS is synthesized by passing a high TC and high spatial coherence (SC) light beam coming from He-Ne laser through a rotating diffuser followed by a MMFB of effective diameter of ~ 5 mm. Linnik interferometer based OCM system is utilized to measure the LSC length of PTS as a function of beam spot size at the diffuser plane. It is observed that the axial resolution (half of LSC length) depends on the beam spot size at diffuser plane and found to be minimum ~ 13 µm for 3.5 mm beam size. The microscope objective with 10× magnification and 0.3 NA is utilized during all experimentation. To further confirm this, interferograms of two standard gauge blocks placed side by side (height difference = 15 µm) are recorded for different beam spot sizes. The change in the visibility of interference pattern formed due to left side gauge block is measured keeping approximately constant visibility at the right side gauge block. The visibility of the interference pattern at left side gauge block is decreased as the beam spot size at the diffuser plane is increased. Thus, high axial resolution and wide field of view (FOV) can be achieved in OCM/T system with high TC ($l_c$ ~ 15 cm) laser. This is contrary to the conventional OCM/T system which utilizes low TC light sources (like halogen lamp, LEDs and broadband lasers etc.) to obtain high axial resolution. The high monochromaticity of the lasers allows the widespread penetration of PTS in OCM/T systems for biological applications.

## 2. Materials and methods
### 2.1. Coherence theory of optical fields

In the coherence theory of optical fields, Wiener–Khintchine theorem is used for the determination of TC function[16,17]. Temporal coherence describes fixed or constant phase relationship, i.e., correlation between light vibrations at two different moments of time. According to this theorem, autocorrelation or temporal coherence function $\Gamma(\Delta t) = \langle E(t)E^*(t - \Delta t) \rangle$ and source power spectral density forms Fourier transform pairs and given by the following relation[14,16].

$$\Gamma(\Delta t) = \int_{-\infty}^{\infty} S(\nu) \exp(i2\pi\nu\Delta t) \, d\nu \quad (1)$$

Where, $\Gamma(\Delta t)$ is the temporal coherence function, $S(\nu)$ is the source spectral distributaion function, and $\Delta t$ is the temporal delay between optical fields $E(t)$ and $E^*(t - \Delta t)$.

The generalized van Cittert–Zernike theorem relates LSC function to the spatial structure (i.e., angular frequency spectrum) of the quasi monochromatic extended light source [14]. Analogous to the Wiener–Khintchine theorem[17,18], the generalized van-Cittert–Zernike theorem [14,16,19] states that LSC function '$\Gamma(\delta z, \Delta t = 0)$' and source angular frequency spectrum form Fourier transform pairs. The LSC function is defined as follows:

$$\Gamma(\delta z, \Delta t = 0) = \int_{-\infty}^{\infty} S(k_z) \exp(i k_z \delta z)\, dk_z, \quad (2)$$

where $\Gamma(\delta z, \Delta t = 0)$ is the LSC function and $S(k_z)$ is the angular frequency spectrum of the light source, $\delta z\ (= z_1 - z_2)$ is the separation between spatial points $Q_1(z_1)$ and $Q_2(z_2)$ situated in two different observation planes along the propagation direction of field, and $k_z$ is the longitudinal spatial frequency[16].

The general expression of the LC length ($L_c$) which depends on both the angular frequency and temporal frequency spectrum of the light source, as follows[16]:

$$L_c = \left[\frac{2\sin^2\left(\theta_z/2\right)}{\lambda} + \frac{\Delta\lambda}{\lambda^2}\cos^2\left(\theta_z/2\right)\right]^{-1} \quad (3)$$

where, $\theta_z$ is half of the angular spectrum width, $\lambda$ is the central wavelength, and $\Delta\lambda$ is related to the temporal spectrum width of the source.

**2.2. Visibility/contrast measurement of the interferogram**

To measure the visibility/contrast of the interferograms Fourier transform algorithm is employed. The 2D intensity distribution of the interferograms is given by the following expression[20,21]:

$$f(x, y) = a(x, y) + b(x, y) \cos[2\pi i(f_x x + f_y y) + \phi(x, y)] \quad (4)$$

where $a(x, y)$ and $b(x, y)$ are the background (DC) and the modulation terms, respectively. Spatially-varying phase $\phi(x, y)$ contains information about the specimen. $f_x$, $f_y$ are the spatial carrier frequencies of interferogram along $x, y$ axes. In practical applications, it is envisaged that $a(x, y)$, $b(x, y)$ and $\phi(x, y)$ are slowly varying functions compared to the variation introduced by the spatial carrier frequencies $f_x$, and $f_y$.

The above intensity modulation can be rewritten in the following form for convenience

$$f(x, y) = a(x, y) + c(x, y) \exp[2\pi i(f_x x + f_y y)] + c^*(x, y) \exp[-2\pi i(f_x x + f_y y)] \quad (5)$$

where

$$c(x, y) = b(x, y) \exp(i\phi(x, y)) \tag{6}$$

The Fourier transform of Eq. 5 is given as follows:

$$F(\xi_x, \xi_y) = A(\xi_x, \xi_y) + C(\xi_x - f_x, \xi_y - f_y) + C^*(\xi_x + f_x, \xi_y + f_y) \tag{7}$$

The term $A(\xi_x, \xi_y)$ is simply a background (DC) term at the origin in the Fourier plane. The term $C(\xi_x - f_x, \xi_y - f_y)$ corresponds to +1 order term contains information about the object and situated at $(+f_x, +f_y)$. Similarly, $C^*(\xi_x + f_x, \xi_y + f_y)$ is −1 order term situated at $(-f_x, -f_y)$ which carry complex conjugate information about of the specimen. After applying Fourier filtering of zero and + 1 order terms present in Eq. 7, the interferogram contrast 'V' can be obtained using the following expression:

$$V = \frac{2 \times \text{maximum}\left(\text{abs}\left(C(\xi_x - f_x, \xi_y - f_y)\right)\right)}{\text{maximum}\left(\text{abs}\left(A(\xi_x, \xi_y)\right)\right)} \tag{8}$$

**2.3. PTS based OCM/T setup**

The experimental scheme of the developed system is based on the principle of non-common-path Linnik based laser-interference microscopy system (Fig. 1a). The laser light beam coming from He-Ne ($l_c$ = 15 cm) goes towards microscope objective MO$_1$ which illuminates the rotating diffuser (RD) with a diverging beam. MO$_1$ is kept on a single axis (x-axis) translation stage to control the size of the beam spot at the diffuser plane. The scattered photons from the RD are directly coupled into a multiple multi-mode fiber bundle (MMFB) placed at ~1 mm distance from the diffuser plane to maximize the number of coupled photons into MMFB. The diameter of MMFB is 5 mm and contains hundreds of fibers (core diameter of each fiber ~ 0.1 mm).

The RD followed by MMFB generates temporally varying speckle field and eventually reduces the speckle contrast significantly. The output port of MMFB acts as a spatially extended purely monochromatic light source namely pseudo thermal light source (PTS). Thus, generates a temporally high and spatially low coherent light source having short LSC length which depends on the spatial extent of the light source. It has been demonstrated that LSC length becomes short for a large source size previously[11]. Here, the maximum source size is decided by the active diameter of the MMFB. At the same time, it also depends on the illumination at the input port of the MMFB. When scattered photons at the output of RD are pumped into a MMFB then it does not or partially couple light into all multimode fibers of the MMFB as depicted in Fig. 1b. Thus, each fiber of MMFB does not contribute to decide the LSC length of the PTS even for the full opening of the active diameter of MMFB. This overall increases the LSC length or poor the axial resolution of PTS based OCM/T system.

Since, the coupling of scattered photons into individual fiber of MMFB depends on the k-vectors present at the output of RD. The number of k-vectors can be increased by changing the beam spot size at the RD plane. The large beam size at RD plane increases the number of scattering sites; thus, generates a wide range of k-vectors. Therefore, there exists sufficient number of scattered photons to match the NA of each individual fiber of MMFB. This leads to the uniform coupling efficiency of the scattered photons into MMFB as illustrated in Fig. 1c. Further, it increases the contribution of large number of sources (fibers of MMFB) to reduce the LSC length of PTS.

The output port of MMFB is attached with the input port of Linnik based interference microscopy system (Fig. 1a). The combination of lenses $L_1$ and $L_2$ relay the source image (fibers of MMFB) at the back focal plane of the microscope objective $MO_2$ (10X, 0.3 NA) to achieve uniform illumination at the sample plane. The beam splitter BS splits the light beam into two; one is directed towards the sample (S) and the other one towards reference mirror ($M_1$). Both light beams reflected back from S (or $M_2$) and $M_1$ recombine and forms interference pattern at the same beam splitter plane, which is projected at the camera plane with the help of $L_3$. The angle of reference mirror $M_1$ controls the angle between the object and the reference beam, i.e., the fringe width of the interferogram. The reference mirror $M_1$ is kept at a particular angle for which high fringe density without aliasing effect is observed at the charge coupled device (CCD) plane. The CCD camera [Lumenera Infinity 2, 1392 × 1024 pixels, pixel size: 4:65×4:65 μm$^2$] is utilized for all interferometric recordings. Subsequently, Fourier transform method[20] is employed for the measurement of visibility/contrast of the recorded interferograms. The image post processing took ~ 1 – 2 s.

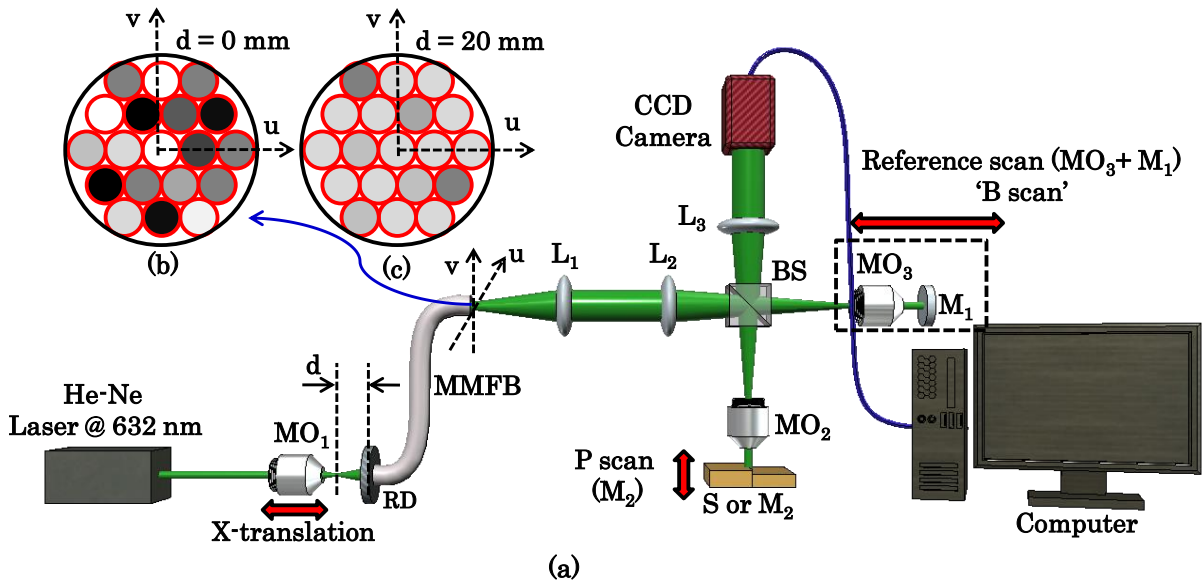

**Fig.1.** (a) Schematic diagram of the spatial coherence gated FF-OCT and QPM system. $MO_{1-3}$: Microscope objectives; BS: Beam splitter; $L_{1-3}$: Lenses; RD: Rotating diffuser; MMFB: Multiple multi-mode fiber bundle; S: sample; $M_{1-2}$: Mirrors and CCD: Charge coupled device. d is the distance between focal position of $MO_1$ and RD plane. (b, c) Intensity distribution at the output of MMFB at d = 0 and 20 mm, respectively. The different gray values represent the amount of intensity at each fiber of MMFB.

## 2.4. Gauge block sample

The standard gauge blocks are utilized to understand the influence of distance 'd' on the LSC length of PTS based OCM/T system. Two gauge blocks of height difference of 15 µm are placed side by side as depicted in Fig. 2a. The height difference between the gauge blocks is confirmed by employing white light interference microscopy (WLIM) system. The details of WLIM can be found in refs.[22,23]. Since, the temporal coherence length of white light is around 1.6 µm, therefore, the height difference of 15 µm between the gauge blocks is confirmed by shifting the white light fringe from the left side gauge block to right side gauge block with an accuracy of 0.8 µm (half of the coherence length of white light). This shifting of white light fringe is done by vertically translating the sample stage in a step of 1 µm. Figures 2b and 2c illustrate the recorded white light interference patterns on the left side and right side gauge block, respectively. The white light fringe width variation on the left and right side gauge blocks could be due to the different angles of gauge blocks as the fringe spacing depends on the angle between object and reference beams. These gauge blocks are further utilized to understand the influence of distance 'd' on the LSC length of PTS.

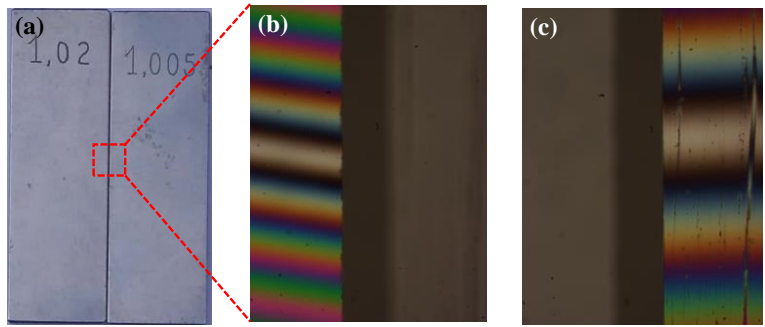

**Fig.2.** (a) Standard gauge blocks of height difference of 15 µm. (b, c) White light interferograms on the left and right side gauge block. The shifting of white light fringe is done by vertically translating the sample stage in a step of 1 µm.

## 3. Results and discussion
### 3.1. Longitudinal spatial coherence measurement

To understand the influence of beam spot size at the RD plane on the LSC length of PTS, $MO_1$ is translated towards or away from RD plane as illustrated in Fig. 1a. A Linnik interference microscopy system is used to realize LSC function experimentally as a function of distance between the focal position of $MO_1$ and RD plane. For the measurement of LSC length or axial resolution (i.e., LSC/2), a flat mirror $M_2$ (Fig. 1a) as a test sample is placed under the interference microscope and scanned vertically in a step of 1 µm from − z to + z (P scan) to sequentially acquire a series of interferograms. The visibility/contrast of the series of interferograms are then measured by employing Eq. 8 and plotted as a function of vertical positions of $M_2$ as depicted in Fig. 3a. The blue solid curve illustrated in Fig. 3a exhibits the LSC function of pseudo thermal light source. It is observed that the fringe visibility of interferograms reduces as $M_2$ go away from the focal position of microscope objective $MO_2$. The FWHM of the LSC function thus obtained provides information about the axial resolution 'Δz' and subsequently LSC length (=2Δz) of PTS.

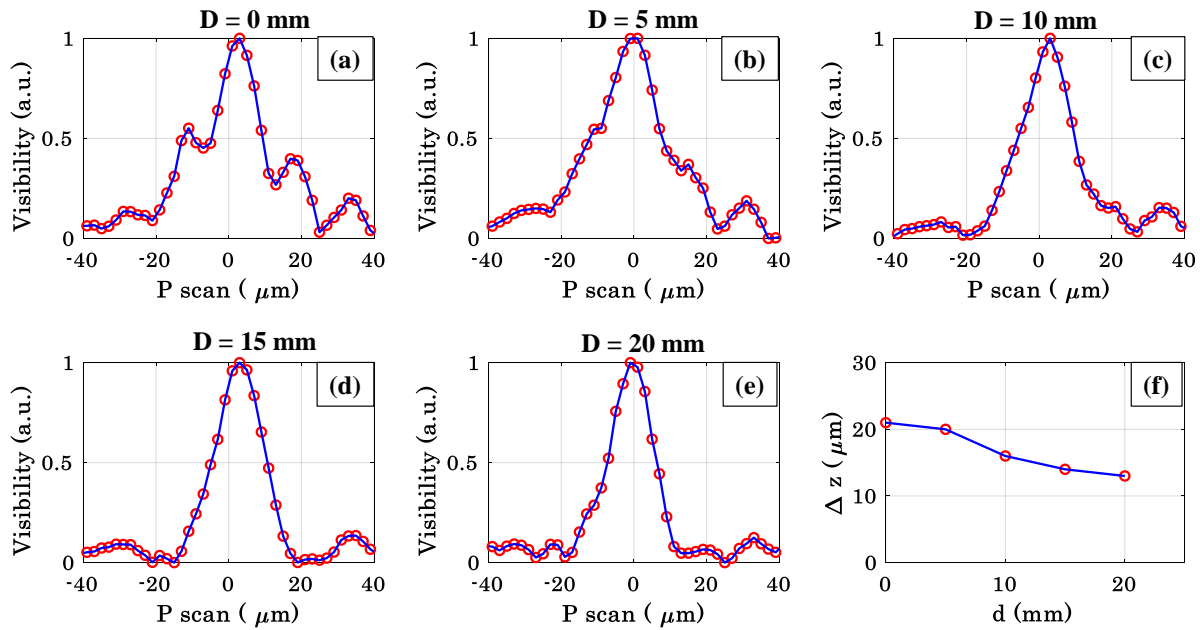

**Fig. 3.** The LC functions of extended monochromatic light source synthesized from a high coherent He-Ne laser ($l_c \sim 15$ cm) as a function of distances 'd' (a – e) 0 – 20 mm in a step of 5 mm, respectively. (f) Variation of the axial resolution of PTS as a function of distance 'd'.

**Table 1.** The axial resolution 'Δz' of the PTS synthesized from He-Ne laser as a function of distance 'd' between focal position of $MO_1$ and RD plane.

| S. No. | 'd' (mm) | Beam spot size at diffuser plane (mm) | Δz (μm) |
|---|---|---|---|
| 1. | 0 | 0.9 | 21 |
| 2. | 5 | 1.5 | 20 |
| 3. | 10 | 2.4 | 16 |
| 4. | 15 | 2.8 | 14 |
| 5. | 20 | 3.5 | 13 |

Figures 3a – 3e present the normalized visibility curves of PTS for the values of d equal to 0, 5, 10, 15, 20 mm, respectively. These values of d correspond to the beam spot sizes of 0.9, 1.5, 2.4, 2.8 and 3.5 mm at the diffuser plane. A slight asymmetry in the measured visibility curves could be due to slight misalignment in the light beam path. The FWHM of each visibility curves are then calculated to obtain axial resolution of the system as given in Table 1. It can be clearly seen from Table 1 that the axial resolution is found to be maximum for the distance of 20 mm. The reason behind obtaining the short LSC length for the large separation of 20 mm between focal plane of $MO_1$ and RD plane has been explained in the materials and method section.

Figure 3f represents the variation of axial resolution as a function of distance between the focal plane of $MO_1$ and RD plane. It is worth noting that axial resolution decreases to 13 μm at the value of 'd' equal to 20 mm. Thus, high axial resolution can be obtained with a monochromatic laser irrespective of its high TC length. The LSC length, i.e., axial resolution, of the system can be further improved with the employment of high NA objective lens as the use of high NA further widen the angular spectrum light source[11,14,24]. As already discussed, the use of such light sources are advantageous as it does not require any dispersion compensation and chromatic aberration corrected optics, which are otherwise mandatory in case of broadband light sources[8].

**3.2. Effect on the fringe visibility of gauge blocks**

To further confirm the reduction in the LSC length of PTS as the distance between the focal position of MO1 and RD plane is increased, experiments are conducted on standard gauge blocks. Two standard gauge blocks having height difference of 15 μm are placed side by side as illustrated in Fig. 2a. Their height difference is confirmed by WLIM as presented in Figs. 2b and 2c. Figures 4a – 4e illustrate the recorded interferograms of the same gauge blocks for d equal to 0, 5, 10, 15, 20 mm, respectively, by employing the experimental setup presented in Fig. 1a. The interferograms of the gauge blocks are recorded keeping approximately constant visibility at the right side gauge block. It is clearly visualized from Figs. 4a – 4e that the fringe contrast is reduced on the left side gauge block as the distance between focal position of MO1 and RD plane is increased. It means that the axial resolution of the PTS based OCM/T system can be increased by appropriately choosing the beam spot size at the RD plane, which is controlled by translating MO1 towards or away from RD plane. The experiments are conducted with 10X (0.3 NA) objective lens for this study.

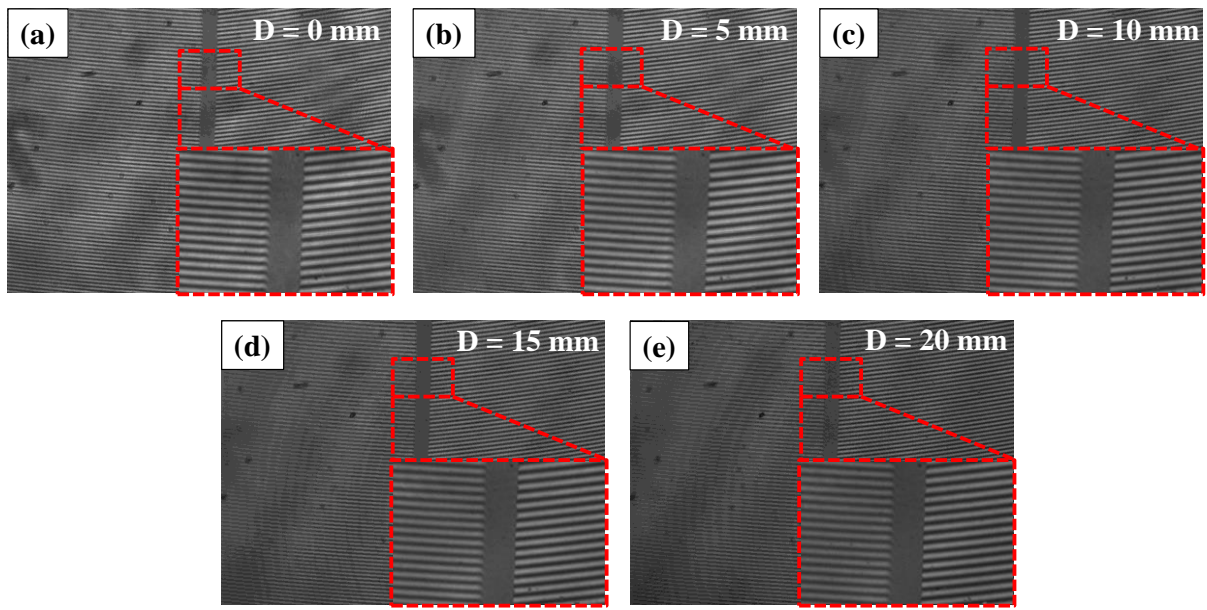

**Fig. 4.** Visibility of the interferograms at the left and the right side gauge block as a function of distance 'd'. (a – e) Recorded interferograms of two standard gauge blocks having 15 μm height difference corresponding to the values of d equal to 0, 5, 10, 15, 20 mm. The visibility is kept constant visibility at the right side gauge block.

Table 2 illustrates the calculated values of the fringe contrast at the left and the right side gauge blocks corresponding to Figs. 4a – 4e. The visibility of the interferograms of the standard gauge blocks are normalized with respect to the right side gauge block. It can be clearly seen that the visibility of the interferogram at the left side gauge block is reduced to 0.44 for the value of d equal to 20 mm.

**Table 2.** Variation of visibility of the interferograms at the left and the right side gauge block as a function of distance 'd'. The visibility at the right side gauge block is kept approximately constant to compare the change in the fringe contrast at the left side gauge block.

| S. No. | d (mm) | Visibility (a.u.) | | Norm. visibility w.r.t. right gauge block (a.u.) | |
|---|---|---|---|---|---|
| | | Left gauge block | Right gauge block | Left gauge block | Right gauge block |
| 1. | 0 | 0.2089 | 0.3058 | 0.6832 | 1 |
| 2. | 5 | 0.1931 | 0.3018 | 0.6399 | 1 |
| 3. | 10 | 0.1787 | 0.3063 | 0.5836 | 1 |
| 4. | 15 | 0.1545 | 0.2991 | 0.5166 | 1 |
| 5. | 20 | 0.1355 | 0.3065 | 0.4421 | 1 |

## 4. Conclusion

In conclusion, high resolution optical sectioning of the samples is possible with PTS based OCM/T system which is otherwise not possible with the direct laser. This is contrary to the principle of conventional OCM/T system, which utilized low TC length sources like halogen lamp or LEDs to perform high resolution sectioning of the biological specimens. The influence of source size at the RD plane by translating the microscope objective MO1 on the axial resolution of PTS is systematically studied. The axial resolution of PTS is measured to be equal to ~ 13 μm for the value of d equal to 20 mm which is otherwise not possible with the monochromatic laser like He-Ne of TC length of 15 cm. Thus, high resolution sectioning of the specimen can be achieved with the employment of a sufficiently wide angular spectrum, i.e., low LSC length, PTS irrespective of the TC length of the parent laser. In addition, any appropriate laser (wavelength falling in the biological window) compatible to the biological samples can be used for the synthesis of PTS and further implemented for high resolution optical sectioning with large penetration depth. Further, PTS generates speckle free images of the biological specimens, i.e., provides high measurement sensitivity, which is otherwise a serious problem in the broadband laser based existing OCM/T systems. In future, PTS will be implemented for performing high axial resolution sectioning of the biological samples with large penetration depth. This type of light source does not require any chromatic aberration corrected optics and dispersion compensation mechanism unlike broadband light source like halogen lamp based OCM/T systems.

**Funding.**

The authors are thankful to Department of Atomic Energy (DAE), Board of Research in Nuclear Sciences (BRNS) for financial grant no. 34/14/07/BRNS.